\documentclass[12pt]{article}

\usepackage{amssymb}
\usepackage{amsmath}
\usepackage{array} % to improve presentation of tables
\usepackage{epsfig}
\usepackage{graphics}

\begin{document}

\begin{center}

%\vskip 7cm
 {\Large \bf
\vskip 7cm
\mbox{Can We Get Deeper}
\\
\mbox{Inside the Pion at the LHC?}
}
\vskip 1cm

\mbox{V.A.~Petrov$^{*}$, R.A~Ryutin$^{*}$, A.E.~Sobol$^{*}$ and M.J.~Murray$^{**}$}
\vskip 1cm

\mbox{{\small $^{*}$Institute for High Energy Physics, {\it 142 281} Protvino, Russia}}

\mbox{{\small $^{**}$University of Kansas, USA}}

 \vskip 1.75cm
{\bf
\mbox{Abstract}}
  \vskip 0.3cm
%---------------Abstract------------------------------------------

\newlength{\qqq}
\settowidth{\qqq}{xxxxxxxxxxxxxxxxxxxxxxxxxxxxxxxxxxxxxxxxxxxxxxxxxxxxxxxxxxxxxxxxxxxxxxxxx}
\hfill
\noindent
\begin{minipage}{\qqq}
We propose a measurement of leading neutrons spectra at 
LHC in order to extract 
inclusive $\pi^+ p$ and $\pi^+\pi^+$ cross-sections 
with high $p_T$ jets production. The cross-sections 
for these processes are simulated with the use of parton distributions 
in hadrons. In this work we estimate 
the possibility to extract parton distributions in the pion from 
the data on these cross-sections 
and also search for 
signatures of fundamental differences 
in the pion and proton structure.

\end{minipage}
\end{center}

%-------------------------------  KEYWORDS  -------------------------------------

\begin{center}
\vskip 0.5cm
{\bf
\mbox{Keywords}}
\vskip 0.3cm

\settowidth{\qqq}{xxxxxxxxxxxxxxxxxxxxxxxxxxxxxxxxxxxxxxxxxxxxxxxxxxxxxxxxxxxxxxxxxxxxxxxxx}
\hfill
\noindent
\begin{minipage}{\qqq}
Leading Neutron Spectra -- Total cross-section -- Absorption -- Regge-eikonal model -- Parton distributions in the pion -- Jets
\end{minipage}

\end{center}

\setcounter{page}{1}
\newpage
\newpage

\section{Introduction}
In recent papers~\cite{ourneutrontot}-\cite{ourneutronregg} we have considered 
the 
possibility 
of useing the LHC as pion-proton and pion-pion collider. Here we
continue to study 
the 
prospect of 
making unique measurements to extract 
cross-sections for $\pi$p and $\pi\pi$ interactions at TeV energies.

Motivations for the present analysis are quite obvious. As one of the 
simplest QCD bound states and as the Goldstone boson of chiral 
symmetry breaking, the pion is a very interesting theoretical object: its 
structure carries important implications for the QCD confinement
mechanism and the realization of symmetries like isospin in nature. It 
is also of practical importance for the hadronic input to the photon 
structure at low scales. The latter is connected via Vector Meson Dominance 
to the meson structure, which is poorly known and thus often 
replaced by the pion structure. 

The parton distributions (PDFs and GPDs) of the nucleons are now 
well determined by global analyses of the precise
data for deep inelastic lepton-nucleon scattering,
Drell-Yan and prompt-photon production. The recent summaries can be found 
in~\cite{durham}-\cite{GPDsreview}, in this work 
we use distributions from~\cite{MSTW} integrated into PYTHIA~\cite{pythia}.  The 
covered region is
$$
10^{-6}<x<1,\; \sim 1\;{\rm GeV}^2<Q^2<10^9\;{\rm GeV}^2.
$$
Unfortunately, determinations of the pion structure have made little progress over 
the last decade (see~\cite{durham} or~\cite{pidistg}-\cite{sumrules} and refs. therein). They 
are based on old Drell-Yan and 
prompt photon data at fixed 
target energies and large values of the partonic momentum fraction $x>0.2$. Many details are 
still based on pure theoretical 
assumptions, partially the precise knowledge of the nucleon distributions, and the use of 
different sum rules~\cite{sumrules} 
that relate nucleon and pion distributions. In order to improve the situation, it has 
been proposed to measure the (virtual) 
pion structure at low $x$ (down to $x\sim 10^{-4}$) in deep inelastic scattering (DIS) 
and photoproduction with leading neutrons 
at HERA~\cite{piDIS1},\cite{piDIS2}. Since the pion is by far the lightest hadron, its 
exchange dominates the $p\to n$ transition 
and it will almost be on its mass shell, particularly at small values of the squared 
momentum transfer $t$ between the proton and 
the neutron. Analysis and references to the DIS dijets data are presented 
in~\cite{piDIS1}-\cite{exp2jet2}. Now we have pion 
distributions in the region
$$
10^{-5}<x<1,\; 5\;{\rm GeV}^2<Q^2<1.31\cdot 10^6\;{\rm GeV}^2.
$$

The parton model of QCD gives us the simple representation of a proton as a three quark state 
and the pion as a quark-antiquark one. This difference 
in the internal structure of nucleons and pions can be investigated experimentally. 
A long 
time ago it was proposed to measure forward-backward 
asymmetry and jet multiplicities in $p p$ and $\pi p$ reactions with high $p_T$ 
events~\cite{fwbkassym}. The assymmetry can serve a clear 
signal that partons have different momentum distributions in protons and pions. It 
is possible to perform such analysis for $\pi\pi$ reaction also.

In this note we consider 
the production of
leading neutrons plus inclusive dijet state, i.e. processes 
of the type 
\begin{equation}
\label{eq1}
pp \to n\ jjX
\end{equation}
and
\begin{equation}
\label{eq2}
pp \to n\ jjX\ n.
\end{equation}
These processes may allow us to extract
parton distributions in the pion in 
an
unprecedentally wide kinematical region:
$$
10^{-6}<x<1,\; \sim 10\;{\rm GeV}^2<Q^2<10^8\;{\rm GeV}^2
$$
for $\sqrt{s}$ up to 14~TeV.

\section{Extraction of $\pi$ p and $\pi$ $\pi$ cross-sections from single and double 
pion exchange measurements.}

\begin{figure}[ht!]
  \hskip 1cm\includegraphics[width=.8\textwidth]{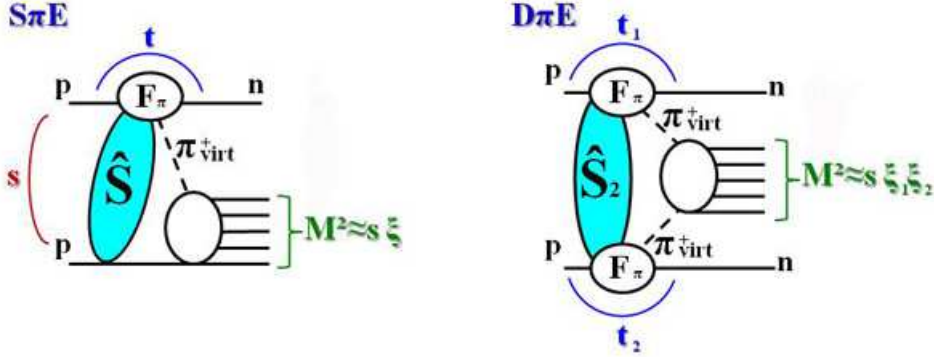}
  \caption{\label{fig:1} Diagrams of Single (S$\pi$E) and Double (D$\pi$E) pion exchanges.}
\end{figure}

  In this section we give an outline of calculations of pion exchange processes 
  with leading neutron production. Diagrams for 
  Single (S$\pi$E) and Double (D$\pi$E) pion exchange processes are presented 
  in Fig.~\ref{fig:1}. Factors $F_{\pi}$ can be 
  normalized to the low energy data~\cite{constG, HERA2} and expressed as
\begin{equation}
\label{eq:1}
F_{\pi}(\xi,t)=
\frac{G^2_{\pi^+pn}}{16\pi^2}\frac{-t}{\left( t-m_{\pi}^2 \right)^2}{\rm e}^{2bt}\xi^{1-2\alpha_{\pi}(t)},
\end{equation}
where
$-t\simeq({\vec{q}}^{\;2}+m_p^2\xi^2)/(1-\xi)$, the pion trajectory is 
$\alpha_{\pi}(t)=\alpha^{\prime}_{\pi}(t-m_{\pi}^2)$ with the slope 
$\alpha^{\prime}_{\pi}\simeq 0.9$~GeV$^{-2}$. $\xi=1-x_L$, where $x_L$ is the fraction of the 
initial proton's longitudinal momentum carried by the 
neutron, and $G_{\pi^0pp}^2/(4\pi)=G_{\pi^+pn}^2/(8\pi)=13.75$~\cite{constG}. From 
recent data~\cite{HERA2},\cite{KMRn1c16}, we expect $b\simeq 0.3\; {\rm GeV}^{-2}$. 

Absorbtive corrections $S$, $S_2$ are estimated in our model for high energy diffractive 
scattering. Details of calculations can 
be found in~\cite{ourneutrontot, ourneutronel}. Final 
formulas for cross-sections look as follows
\begin{eqnarray}
\label{csSpiE} \frac{d\sigma_{{\rm S}\pi{\rm E}}}{dt\;d\xi}&\!\!\!\!=&\!\!\!\! 
F_{\pi}(\xi,t)S\left( s/s_0,\xi,t \right)\sigma_{\pi p}(s\;\xi),\\
\label{csDpiE} \frac{d\sigma_{{\rm D}\pi{\rm E}}}{dt_1 dt_2 d\xi_1 d\xi_2}&\!\!\!\!=&\!\!\!\! 
F_{\pi}(\xi_1,t_1)F_{\pi}(\xi_2,t_2) S_2\left( s/s_0,\xi_{1,2},t_{1,2} \right)
\sigma_{\pi\pi}(s\;\xi_1\xi_2).
\end{eqnarray}
Here 
$$
\sigma_{\pi p}(\hat{s};\{ m_p^2, m_{\pi}^2\})\simeq\sigma_{\pi_{virt} p}(\hat{s};\{ m_p^2, t\}), 
\sigma_{\pi\pi}(\hat{s};\{ m_{\pi}^2\})\simeq\sigma_{\pi_{virt}\pi_{virt}}(\hat{s};\{ t_{1,2}\}),
$$
since the main contribution comes from pions with very low virtualities $|t_i|<0.3\;{\rm GeV}^2$. We 
are interested in the kinematical range 
$0.01$~GeV$^2<|t_i|<0.5$~GeV$^2$, $\xi_i<0.4$, where 
formulaes~(\ref{csSpiE}),(\ref{csDpiE}) dominate according to~\cite{KMRn1c13, KMRn1c14}. 

The next question is 
how to extract $\pi$ p and $\pi$ $\pi$ cross-sections from the data on S$\pi$E and D$\pi$E? The 
exact procedure is similar to the Goebel~\cite{goebel} and Chew-Low~\cite{chewlow} method:
\begin{eqnarray}
\label{extrPIPex} \sigma_{\pi p}(s\;\xi)&\!\!\!\!=&\!\!\!\!\lim_{t\to m_{\pi}^2} 
\sigma_{\pi_{virt} p}(s\;\xi;\{ t\}) \frac{S\left( s/s_0,\xi,t \right)t}{m_{\pi}^2}=
\lim_{t\to m_{\pi}^2} E(\xi,t) \frac{d\sigma_{{\rm S}\pi{\rm E}}}{dt\; d\xi}\\
 \sigma_{\pi\pi}(s\;\xi_1\xi_2)&\!\!\!\!=&\!\!\!\!\lim_{t_{1,2}\to m_{\pi}^2} 
 \sigma_{\pi_{virt}\pi_{virt}}(s\;\xi_1\xi_2;\{ t_{1,2}\}) 
 \frac{S_2\left( s/s_0,\xi_{1,2},t_{1,2} \right)t_1 t_2}{m_{\pi}^4}=\nonumber\\
\label{extrPIPIex} &\!\!\!\!=&\!\!\!\!\lim_{t_{1,2}\to m_{\pi}^2} E(\xi_1,t_1)E(\xi_2,t_2) 
\frac{d\sigma_{{\rm D}\pi{\rm E}}}{dt_1 dt_2 d\xi_1 d\xi_2},\\
\label{extrE} E(\xi,t)&\!\!\!\!=&\!\!\!\!-\frac{\left( t-m_{\pi}^2\right)^2}{m_{\pi}^2} 
\frac{16\pi^2}{G^2_{\pi^+pn}{\rm e}^{2bt}\xi^{1-2\alpha_{\pi}(t)}}.
\end{eqnarray}

\begin{figure}[b!]
  \hskip 1cm\includegraphics[width=.7\textwidth]{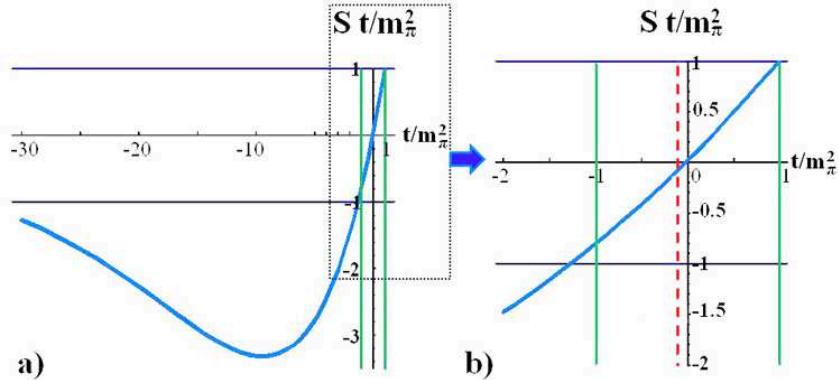}
  \caption{\label{fig:2}Function from the expression~\eqref{extrPIPex} at fixed $\xi=0.05$. The 
  boundary of the physical region $t\simeq-m_p^2\xi^2/(1-\xi)$ is represented by vertical dashed line.}
\end{figure}
The behavior of corresponding functions is shown in the Fig.~\ref{fig:2}. When t is equal to 
mass of the pion squared, we have no absorbtion at all ($S=1$) and 
extracted cross-sections are independent on the model for rescattering corrections.

\begin{figure}[t!]
  \hskip 1cm\includegraphics[width=.7\textwidth]{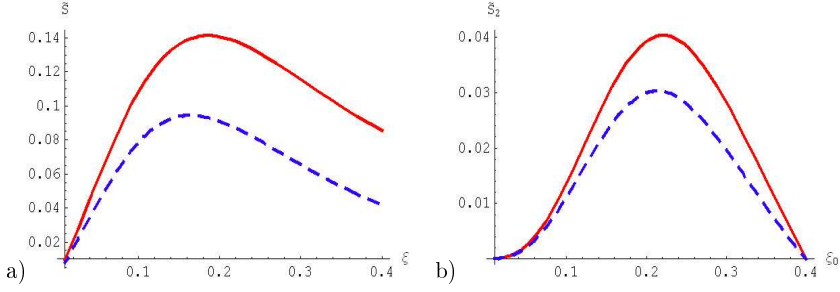}
  \caption{\label{fig:3}Rescattering corrections integrated with formfactors 
  for $\sqrt{s}=0.9$~TeV (solid) and $\sqrt{s}=7$~TeV (dashed): a) $\tilde{S}(s,\;\xi)$; 
  b) $\tilde{S}_2(s,\;\xi_0)$.}
\end{figure}
The real situation is more complicated, especially from 
the experimental point of view. It is rather difficult to measure transverse momentum 
of a fast leading neutron, we can only get some restrictions on $t$ from the acceptance 
of detectors. We propose to use the model dependent integrated method
presented by formulas
\begin{eqnarray}
\label{extrPIPint} \tilde{S}(s,\xi)&\!\!\!\!=&\!\!\!\!
\int_{t_{min}}^{t_{max}}dt\; S\left( \frac{s}{s_0},\xi,t \right)
F_{\pi}(\xi,t),\; \sigma_{\pi p}\left( M_{\pi p}^2\right)=
\frac{\frac{d\sigma_{{\rm S}\pi{\rm E}}}{d\xi}}{\tilde{S}(s,\xi)},\; 
\xi\simeq \frac{M_{\pi p}^2}{s},\\
\label{extrPIPIintA} \tilde{S}_2(s,\xi_0)&\!\!\!\!=&\!\!\!\!
\int_{-y_0}^{y_0}dy\int_{t_{min}}^{t_{max}}dt_1 dt_2\; 
S_2\left( \frac{s}{s_0},\xi_0{\rm e}^{\pm y},t_{1,2} \right)
F_{\pi}(\xi_0{\rm e}^{y},t_1)F_{\pi}(\xi_0{\rm e}^{-y},t_2),\\
\label{extrPIPIintB} \sigma_{\pi\pi}\left( M_{\pi\pi}^2\right)&\!\!\!\!=&\!\!\!\!
\frac{\frac{d\sigma_{{\rm D}\pi{\rm E}}}{d\xi_0}}{\tilde{S}_2(s,\xi_0)},\; 
\xi_0\simeq \frac{M_{\pi\pi}}{\sqrt{s}},\; y_0=\ln\frac{\xi_{max}}{\xi_0}.
\end{eqnarray}
Functions $\tilde{S}$ and $\tilde{S}_2$ are shown in the Fig.~\ref{fig:3}. Models for rescattering 
give us theoretical errors. If we have the data on p p and anti-p p total and
elastic cross-sections, these uncertainties could be reduced to the errors of the 
data. For example, without LHC measurements at 10 TeV theoretical uncertainties can 
be estimated only from model predictions and can reach 20\% for the most popular 
models. These errors are low for energies less than 1.9~TeV, since we have precise 
measurements from Tevatron. 
 
 Our method~\eqref{extrPIPint}
which used a
very narrow t interval was applied to the extraction 
of $\pi^+$ p total 
cross-sections at low energies~\cite{ourneutrontot}. It was shown in~\cite{ourneutrontot} 
that extracted points are 
close to the real data and four different model predictions which is a clear signal of the 
validity of our method.

\section{Pion exchanges with dijet production}

Since it is possible to extract $\pi$ p and $\pi$ $\pi$ 
cross-sections (total, elastic, Drell-Yan, direct photon or inclusive dijet 
production and so on) from the LHC data 
it should be possible to use such results to look
inside the pion as we usually do it with proton 
and anti-proton. In this article we consider only the case of the inclusive dijet 
production as an example.

\begin{figure}[t!]
  \hskip 2cm\includegraphics[width=.6\textwidth]{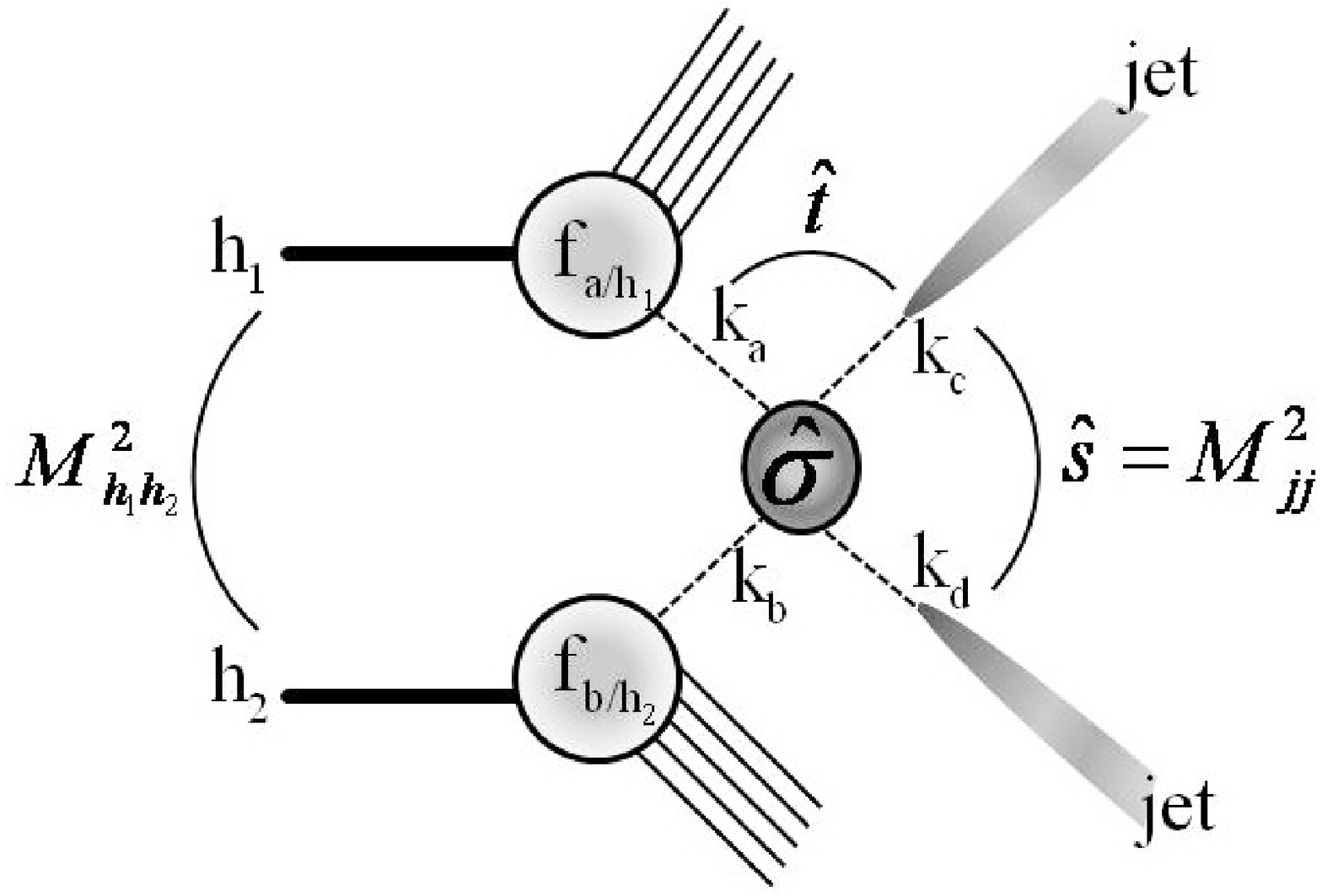}
  \caption{\label{fig:4}Process $h_1+h_2\to{\rm jet}\;{\rm jet}\;{\rm X}$.}
\end{figure}

Let us consider $\pi$ p or $\pi$ $\pi$ scattering with dijet production 
as a general process of the type $h_1+h_2\to {\rm jet}\; {\rm jet}\; {\rm X}$ (see 
Fig.~\ref{fig:4} for
definitions). Momenta of particles can be represented as usual (for 
any $p\equiv\{ p_0,p_3;\;\vec{p}\}$):
\begin{eqnarray}
\label{jjkin1} k_a &\!\!\!\!=&\!\!\!\!\left\{ 
x_a p_{h_1,0}, x_a p_{h_1,3};\; \vec{k}_{t,a}\right\},\; k_b=
\left\{ x_b p_{h_2,0}, x_b p_{h_2,3};\; \vec{k}_{t,b}\right\},\\
\label{jjkin2} k_i&\!\!\!\!=&\!\!\!\!
\left( k_{t,i}\cosh\eta_i, k_{t,i}\sinh\eta_i;\;  \vec{k}_{t,i}\right),
\; i=c,d,\\
\label{jjkin3} \hat{s}&\!\!\!\!\simeq &\!\!\!\! x_a x_b M_{h_1h_2}^2
\simeq M_{jj}^2=(k_c+k_d)^2=(k_a+k_b)^2,\\
\hat{t}&\!\!\!\!=&\!\!\!\!(k_a-k_c)^2=(k_b-k_d)^2.
\end{eqnarray}
In the case of $\pi$ p scattering we have
$$
p_{h_1}=p_{\pi}\simeq\left\{ 
\xi\frac{\sqrt{s}}{2},\xi\frac{\sqrt{s}}{2};\; \vec{q}\right\},\; 
p_{h_2}=p_2\simeq \left\{ \frac{\sqrt{s}}{2},-\frac{\sqrt{s}}{2};\; \vec{0}\right\}
$$
and for $\pi$ $\pi$ scattering
$$
p_{h_1}=p_{\pi,1}\simeq\left\{ \xi_1\frac{\sqrt{s}}{2},\xi_1\frac{\sqrt{s}}{2};\; \vec{q}_1\right\},\; 
p_{h_2}=p_{\pi,2}\simeq \left\{ \xi_2\frac{\sqrt{s}}{2},-\xi_2\frac{\sqrt{s}}{2};\; \vec{q}_2\right\}.
$$
For $|t_i|\ll m_{h_1,h_2}^2\ll M^2_{h_1h_2}$ we have 
\begin{eqnarray}
&&  x_a=\frac{1}{2p_{h_1,0}}\left( m_{\perp,c}\;{\rm e}^{\eta_c}+m_{\perp,d}\;{\rm e}^{\eta_d}\right),\; 
x_b=\frac{1}{2p_{h_2,0}}\left( m_{\perp,c}\;{\rm e}^{-\eta_c}+m_{\perp,d}\;{\rm e}^{-\eta_d}\right),
\nonumber\\ 
&& \label{jjkinxab} m_{\perp,i}=\sqrt{m_i^2+k_{t,i}^2}.
\end{eqnarray}
And in the collinear approximation 
$k_{t,\; a,b}\ll k_t\simeq k_{t,c}\simeq k_{t,d}$, $m_{a,b,c,d}\ll k_t$
\begin{eqnarray}
&& x_a=\frac{k_t}{p_{h_1,0}}{\rm e}^y\cosh\eta,\; x_b=
\frac{k_t}{p_{h_2,0}}{\rm e}^{-y}\cosh\eta,\nonumber\\
&& M_{jj}\simeq 2k_t\cosh\eta,\; \eta=\frac{\eta_d-\eta_c}{2},\; y=
\frac{\eta_d+\eta_c}{2},\nonumber\\
\label{collaprox} && \hat{t}\simeq-\frac{\hat{s}}{2}(1+z),\; \hat{u}\simeq-\frac{\hat{s}}{2}(1-z),\; 
z=\tanh\eta=\cos\theta^*, 
\end{eqnarray}
where $\theta^*$ is the scattering angle in the CM frame of partons a and b.

The basic formula for inclusive two parton production in the collinear approximation looks as follows
\begin{eqnarray}
&& \frac{d\sigma_{h_1+h_2\to c\;d\;X}(M_{h_1h_2}^2)}{dx_a dx_b dz}=
\frac{\hat{s}}{2}\sum_{a,b}f_{a/h_1}(x_a)f_{b/h_2}(x_b)
\frac{d\hat{\sigma}_{ab\to cd}}{d\hat{t}}(\hat{s};\; z)=\nonumber\\
\label{colljjcs} && =\sum_{a,b}f_{a/h_1}(x_a)f_{b/h_2}(x_b)
\frac{d\hat{\sigma}_{ab\to cd}}{dz}(\hat{s};\; z),
\end{eqnarray}
where $f_{i/h}(x)$ is the number density of parton $i$ (quark, anti-quark or gluon) 
with the longitudinal momentum $x\; p_{h,0}$ in the hadron $h$. Renormalization 
and factorization scales are hidden in $G$ and $\hat{\sigma}$. We can reconstruct 
momenta of final partons from jets measurements and then use our 
method~\eqref{extrPIPint}-\eqref{extrPIPIintB} to obtain combinations 
of PDFs in the pion and a proton. Since
we know proton PDFs from other experiments, we can extract combinations of pion's PDFs
\begin{eqnarray}
\label{collextrPIP} && \sum_{a,b}f_{a/\pi}(x_a)\tilde{f}_{b/p}(s\;\xi\; x_a)=\frac{\frac{d\sigma_{{\rm S}\pi{\rm E}}^{jj}}{d\xi dx_a}}{\tilde{S}(s,\xi)},\\
\label{tildef} && \tilde{f}_{b/p}(s\;\xi\; x_a)=\int dx_b\; f_{b/p}(x_b)\hat{\sigma}_{ab\to cd}(s\;\xi\;x_a x_b),\\
\label{collextrPIPI} && \sum_{a,b}f_{a/\pi}(x_a)f_{b/\pi}(x_b)\hat{\sigma}_{ab\to cd}(s\;\xi_0^2 x_a x_b)=\frac{\frac{d\sigma_{{\rm D}\pi{\rm E}}^{jj}}{d\xi_0 dx_a dx_b}}{\tilde{S}_2(s,\xi_0)},
\end{eqnarray}
and $d\hat{\sigma}_{ab\to cd}/dz$ can be found in the Table~\ref{tab:dsigdz}.

\begin{table}[bt!]
\caption{\label{tab:dsigdz} Leading order parton-parton cross-sections 
$d\hat{\sigma}_{ab\to cd}/dz=\frac{\pi\alpha_s^2}{2\hat{s}}A_{ab\to cd}(\hat{s},\hat{t},\hat{u})$.}
\begin{center}
\begin{tabular}{|c|c|}
\hline
 Subrocess  &  $A_{ab\to cd}$ \\
\hline
 $qq'\to qq'$ & $\frac{4}{9}\frac{\hat{s}^2+\hat{u}^2}{\hat{t}^2}$\\
\hline
 $qq\to qq$ & 
 $\frac{4}{9}\left[ 
 \frac{\hat{s}^2+\hat{u}^2}{\hat{t}^2}+
 \frac{\hat{s}^2+\hat{t}^2}{\hat{u}^2}\right]-
 \frac{8}{27}\frac{\hat{s}^2}{\hat{t}\hat{u}}$\\
\hline
  $q\bar{q}\to q'\bar{q}'$ & $\frac{4}{9}\frac{\hat{t}^2+\hat{u}^2}{\hat{s}^2}$\\
\hline
  $q\bar{q}\to q\bar{q}$ & 
  $\frac{4}{9}\left[ \frac{\hat{s}^2+\hat{u}^2}{\hat{t}^2}+
  \frac{\hat{u}^2+\hat{t}^2}{\hat{s}^2}\right]-
  \frac{8}{27}\frac{\hat{u}^2}{\hat{s}\hat{t}}$\\
\hline
  $gq\to gq$ & 
  $-\frac{4}{9}\left[ \frac{\hat{s}}{\hat{u}}+
  \frac{\hat{u}}{\hat{s}}\right]+
  \frac{4}{9}\frac{\hat{s}^2+\hat{u}^2}{\hat{t}^2}$ \\  
\hline
  $q\bar{q}\to gg$ & 
  $\frac{32}{27}\left[ \frac{\hat{t}}{\hat{u}}+
  \frac{\hat{u}}{\hat{t}}\right]-
  \frac{8}{3}\frac{\hat{t}^2+\hat{u}^2}{\hat{s}^2}$ \\
\hline
  $gg\to q\bar{q}$ & 
  $\frac{1}{6}\left[ \frac{\hat{t}}{\hat{u}}+
  \frac{\hat{u}}{\hat{t}}\right]-
  \frac{3}{8}\frac{\hat{t}^2+\hat{u}^2}{\hat{s}^2}$\\
\hline
  $gg\to gg$ & 
  $\frac{9}{2}\left[ 3-\frac{\hat{t}\hat{u}}{\hat{s}^2}-
  \frac{\hat{s}\hat{u}}{\hat{t}^2}-
  \frac{\hat{s}\hat{t}}{\hat{u}^2}\right]$\\  
\hline
\end{tabular}
\end{center}
\end{table}

\section{Experimental possibilities.}
\label{section:exp}

\begin{figure}[h!]
\begin{center}
  \includegraphics[width=\textwidth]{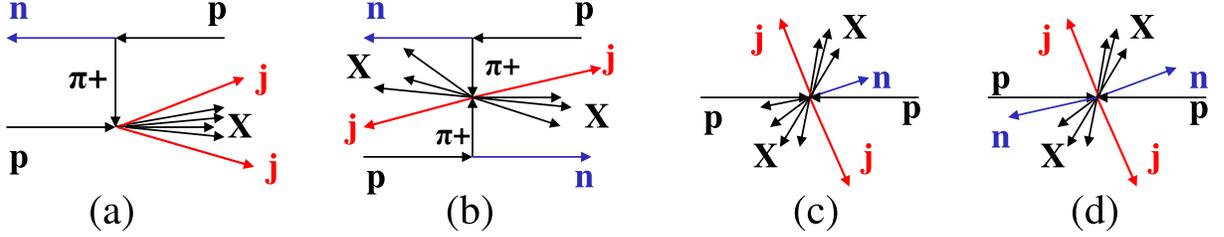}
\end{center}  
  \caption{\label{fig:5} (a) and (b): diagrams for the processes of S$\pi$E and D$\pi$E with 
  2-jet production. 
  (c) and (d): diagrams for the processes of $pp$ inelastic interactions with 2 jets 
  and leading neutrons production, 
  which can imitate processes (a) and (b).}  
\end{figure}

We propose to measure 
jet production in 
single and double pion exchange reactions ((\ref{eq1}), (\ref{eq2})) at the LHC using
the CMS detector~\cite{CMS}. Diagrams of the reactions
 (\ref{eq1}) and (\ref{eq2}) are 
shown on the Fig.~\ref{fig:5} (a) and (b) correspondingly. 

In these processes two jets are produced in the hard $\pi^+p$ and $\pi^+\pi^+$ scattering which allows us
to study parton distributions in pions in
a still unexplored kinematical 
region. In this chapter we discuss perspectives of such measurements at 7 TeV. 
The 
Zero Degree Calorimeters (ZDCs)~\cite{ZDC} can be used~\cite{ourneutrontot, ourneutronel} 
to measure the leading 
neutrons. The ZDCs are placed on the both sides of CMS, 140m away from the interaction point. 
They have electromagnetic and hadronic sections designed to measure photons and neutrons 
in the pseudorapidity region $|\eta|>8.5$.

The 
Monte-Carlo generator MONCHER v.1.0~\cite{monchermanual} has been used for numerical 
simulation of the processes 
(\ref{eq1}) and (\ref{eq2}). This generator is developed by two of the authors of the article 
for S$\pi$E and D$\pi$E simulation 
specially.  The kinematics of S$\pi$E and D$\pi$E reactions are defined by the relative 
energy loss $\xi_n$  and the square 
of the transverse momentum $t_n$ of the leading neutron. The vertex $p\pi^+_{virt}n$ 
is generated according to the 
model described in the Ref.~\cite{ourneutrontot, ourneutronel}. PYTHIA 6.420~\cite{pythia} is 
used for the 
$\pi^+_{virt}p \to X$ generation for the
single pion exchange and $\pi^+_{virt}\pi^+_{virt} \to X$ 
generation
for the double pion exchange. 
All background processes have been generated 
by PYTHIA 6.420. Diagrams of two background processes, $pp$ inelastic interactions with 
2 jets and leading neutrons production 
imitating signal, are presented on the Fig.~\ref{fig:5} (c) and (d).  

PYTHIA 6.420 predicts 90.76~mb for the $pp$ total cross section at 7~TeV. The 
inelastic part 
of this cross section, which is interesting for us as background, consists of 48.4~mb of minimum bias 
events, 13.7~mb of single diffractive events and 9.3 mb of double diffractive events. 
MONCHER 1.0 predicts 1.31$\div$1.85~mb for S$\pi$E and 0.17$\div$0.30~mb for D$\pi$E processes at 
7 TeV\footnote{Cross sections for S$\pi$E and D$\pi$E are given for $\xi_n<0.4$. This is a 
kinematical bound of the model, see Ref.~\cite{ourneutrontot}.}. Uncertainty in the cross 
sections comes from the different models for $\pi^+p$ and $\pi^+\pi^+$ interactions. In this 
study we use the most pessimistic estimations from the Donnachie-Landshoff 
parametrization~\cite{lanshofftot}. Therefore 
before any selections the ratio of the signal and background processes at 7 TeV looks as follows:
$$\rm \sigma_{D\pi E}:\sigma_{S\pi E}:\sigma_{TOTAL} = 1 : 7.6 : 530.$$
In the rest of this paper we seek effective criteria to raise the signal/background ratio for single and double pion exchange. 

\begin{figure}[h!]
\begin{center}
  \includegraphics[width=\textwidth]{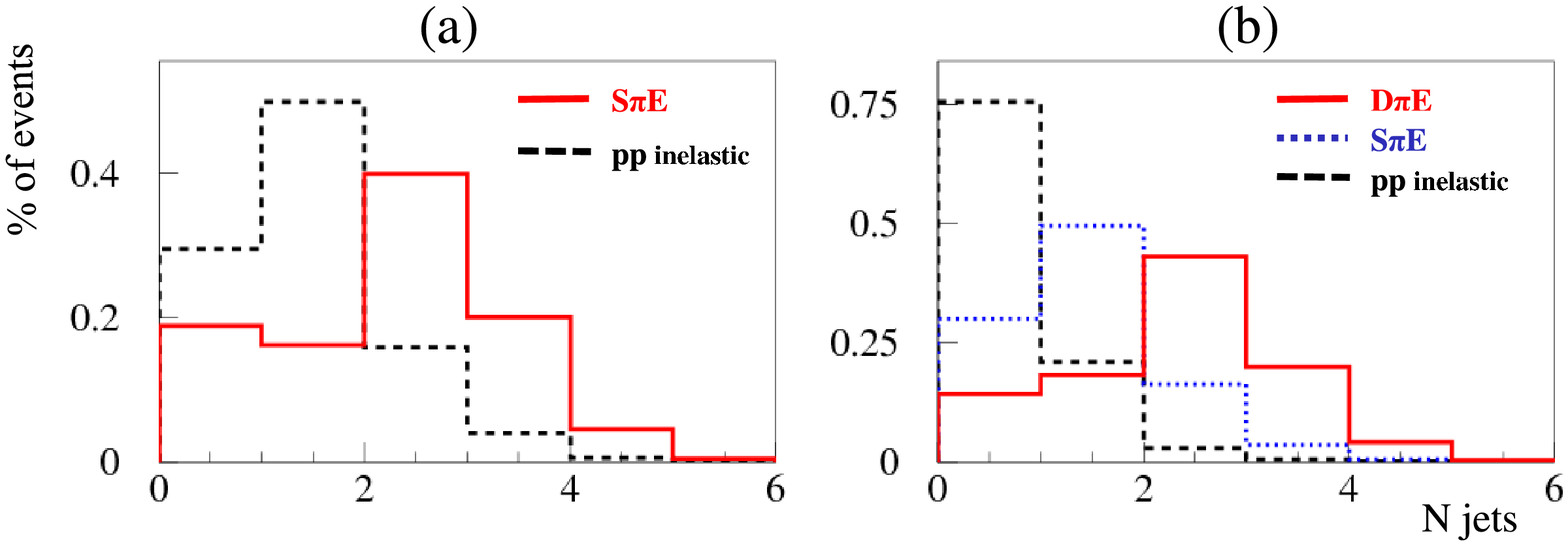}
\end{center}  
  \caption{\label{fig:6} (a) Multiplicity distribution of jets  from the  S$\pi$E (red(solid)) and $pp$ inelastic 
 events (black(dashed)) selected by~(\ref{exp:sel1}).
 (b) Multiplicity distribution of jets  from the  D$\pi$E (red(solid)), S$\pi$E (blue(dotted)) and $pp$ inelastic 
 events (black(dashed)) selected by~(\ref{exp:sel2}).}  
\end{figure}

For S$\pi$E we selected events with signal from neutrons in the forward or backward 
ZDC and with the absence of neutrons in the opposite one:
\begin{equation}
\label{exp:sel1}
\left[
\begin{array}{l} 
\rm N_n^f>0\quad  \& \quad N_n^b=0\quad \&\quad \xi_n^f<0.4 \\
\rm N_n^b>0\quad  \& \quad N_n^f=0\quad \&\quad \xi_n^b<0.4.
\end{array}
\right.
\end{equation}
For the D$\pi$E we selected events with neutrons on both sides:
\begin{equation}
\label{exp:sel2}
\begin{array}{l} 
\rm N_n^f>0\quad  \& \quad N_n^b>0\quad  \&\quad  \xi_n^f<0.4\quad  \& \quad \xi_n^b<0.4.
\end{array}
\end{equation} 

Here, $N_n^f$ ($N_n^b$) is the number of neutrons hitting 
the forward (backward) ZDC, $\xi_n^f$ ($\xi_n^b$) is the relative energy loss 
of the forward (backward) neutron. 
The signal to background ratio becomes equal to 0.22 for the  S$\pi$E and 0.76 
for the D$\pi$E after selections 
(\ref{exp:sel1}) and (\ref{exp:sel2}) respectively. For S$\pi$E all background 
events are produced in the inelastic $pp$ interactions (minimum bias, single and double diffraction). 
For  D$\pi$E 20\% of background are imitated by S$\pi$E and 80\% come from $pp$ inelastic interactions. 

Fig.\ref{fig:6} presents distributions of the  multiplicity $N^{jets}$ of the jets 
from  S$\pi$E,  D$\pi$E and $pp$ inelastic events selected by (\ref{exp:sel1}) (a) and by 
(\ref{exp:sel2}) (b). Signal (red solid histograms) and background (blue dotted and black dashed) 
have rather different distributions of $N^{jets}$. For the analysis we selected 2-jet events 
dominating  in the  S$\pi$E and D$\pi$E production(around 40\%). 

\begin{figure}[h!]
\begin{center}
  \includegraphics[width=\textwidth]{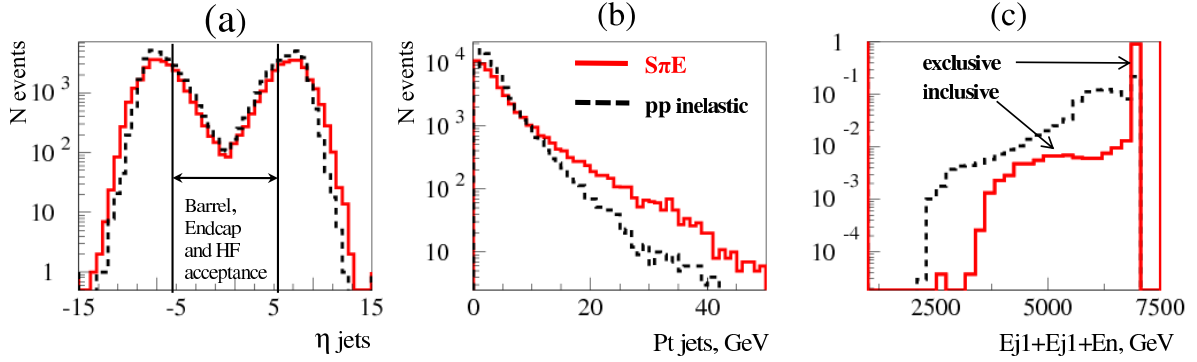}
\end{center}  
  \caption{\label{fig:7} Distributions in $\eta^{jets}$ (a), transverse momentum $p_t^{jets}$ (b) 
 of jets and the sum of jets and neutron energies,$E_{jjn}=E_{j1}+E_{j2}+E_{n}$, from the  
 2-jet S$\pi$E (red(solid)) and $pp$ inelastic events (black(dashed)) selected by~(\ref{exp:sel1}).
 The last right bin of the distribution (c), peaking at 7 TeV, corresponds to the exclusive 
 production of jets. Events with  $E_{jjn}<7$ TeV on the plot (c) come from the inclusive 
 jets production. Vertical lines on the plot (a) shows Barrel, Endcap and HF acceptance of the CMS.}   
\end{figure}

Fig. \ref{fig:7} (a) and (b) shows pseudorapidity distributions in $\eta^{jets}$ and transverse momentum $p_t^{jets}$  
of jets from the 2-jet S$\pi$E (red(solid)) and $pp$ inelastic (black(dashed)) events
selected by~(\ref{exp:sel1}). 
The 
$p_t^{jets}$ distribution of the jets from signal (red (solid) histogram) 
shows 
a
more gentle sloping  behaviour comparing with jets from background (black (dashed) histogram). 
It can be used for the further signal/background separation. 
Vertical lines on the plot (a) show  the total Barrel, Endcap and HF acceptance of CMS.
Plot (c) on the Fig.~\ref{fig:7} presents the distribution of the sum of jets and 
neutron energies, $E_{jjn}=E_{j1}+E_{j2}+E_{n}$, for the signal and background 2-jet events 
selected by~(\ref{exp:sel1}).
The last right bin of the distribution, peaking at 7~TeV, corresponds to the exclusive 
production of jets. Events with  $E_{jjn}<7$~TeV  come from the inclusive 
jets production.  It is seen that 2-jet signal events are produced in the exclusive process dominantly
(ratio of exclusive to inclusive production is approximately equal to $8.4$).  
For the background, inversely, inclusive production of 2-jet events is more intensive 
(exlusive/inclusive ratio is approximately equal to $0.23$). We also use this difference for the further
signal/background separation.
\begin{figure}[h!]
\begin{center}
  \includegraphics[width=\textwidth]{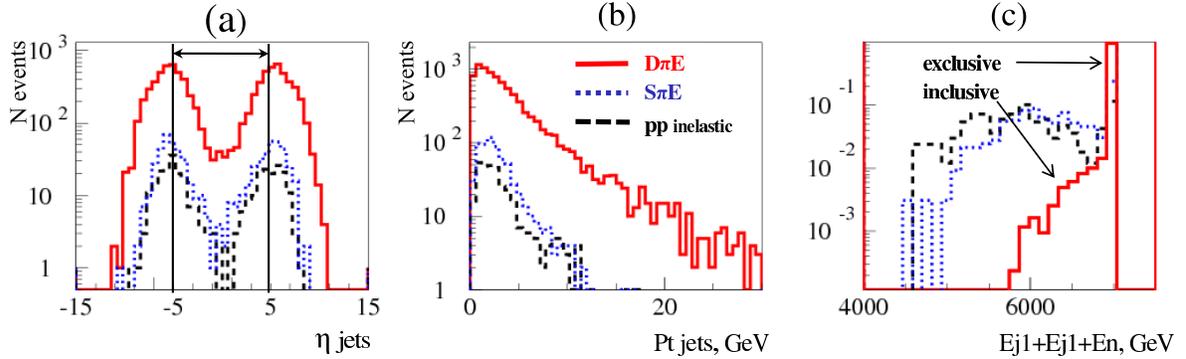}
\end{center}  
  \caption{\label{fig:8} The same as on the Fig.\ref{fig:7} for the  2-jet D$\pi$E (red(solid)), 
  S$\pi$E (blue(dotted)) and $pp$ inelastic events (black(dashed)) selected by~(\ref{exp:sel2}).}
\end{figure}

Fig.~\ref{fig:8} shows 
the same distributions as Fig.~\ref{fig:7} for
D$\pi$E (red (solid) histogram),  S$\pi$E (blue (dotted)) and $pp$ inelastic (black (dashed)) 2-jet 
events selected  by (\ref{exp:sel2}). The difference between signal and background in the $p_t^{jets}$ 
distributions becomes more essential. Practically, there are no background events at  
$p_t^{jets}>10$ GeV, which can be used for the total separation of the 2-jet D$\pi$E events from
background. Almost all  2-jet D$\pi$E events selected by~(\ref{exp:sel1})  are produced 
exclusively. The  
ratio of exclusive to inclusive production is equal $\approx 14$.

Analysing distributions of jets for signal and background we suggest 
the following cuts 
\begin{equation}
\label{exp:sel3}
\left\lbrace
\begin{array}{l} 
\rm |\eta^{jets}|<5 \\
\rm p_t^{jets}>30\ GeV, \\
\end{array}
\right.
\end{equation}
for 2-jet S$\pi$E events selection and 
\begin{equation}
\label{exp:sel4}
\left\lbrace
\begin{array}{l} 
\rm |\eta^{jets}|<5 \\
\rm p_t^{jets}>10\ GeV, \\
\end{array}
\right.
\end{equation}
for 2-jet D$\pi$E events selection. $p_t^{jets}$-cut can be varied depending on trigger requirements and 
number of detected events to optimize signal/background ratio. 

Events of the reaction~(\ref{eq1}) selected by (\ref{exp:sel1})\&(\ref{exp:sel3}) have $\approx$15\% 
of background
from 2-jet $pp$ inelastic production with leading neutrons. 
Events of the reaction~(\ref{eq2}) selected by (\ref{exp:sel2})\&(\ref{exp:sel4}) have $\approx$3\%  
of background from $pp$ inelastic and S$\pi$E production imitating signal.
The additinal requirement
\begin{equation}
\label{exp:sel5}
\rm 6990 < \sum{(E_{jets}+E_{neutrons})} < 7010\ GeV,
\end{equation}
selecting events with exclusive jets production, allows to suppress background for~(\ref{eq1}) down 
to the level 
6.5\% and completely suppress the background for~(\ref{eq2}).
\begin{figure}[h!]
\begin{center}
  \includegraphics[width=0.9\textwidth]{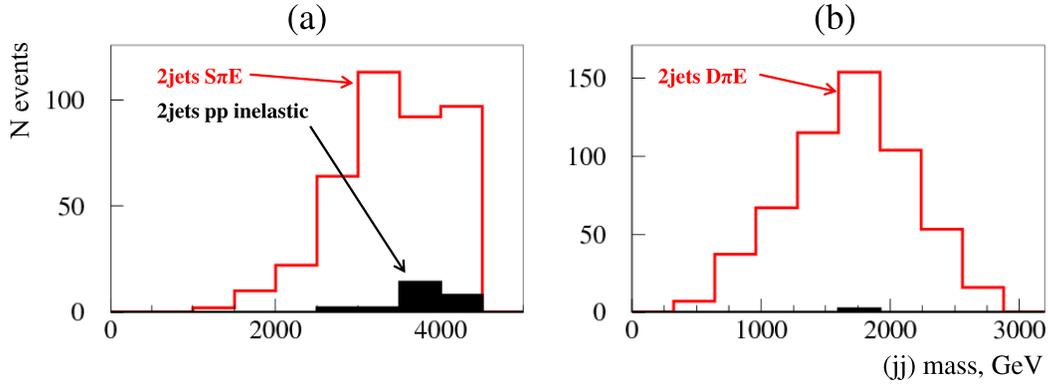}
\end{center}  
  \caption{\label{fig:9} Distributions of events in (a) Invariant mass of (jj) system  for S$\pi$E events selected by 
   (\ref{exp:sel1})\&(\ref{exp:sel3})\&(\ref{exp:sel5}) is shown by red (solid) line. 
   3\% of background from $pp$ inelastic events is shown by black histogram.
   (b)  Invariant mass of (jj) system  for D$\pi$E events selected by 
   (\ref{exp:sel2})\&(\ref{exp:sel4})\&(\ref{exp:sel5}).}  
\end{figure}

Invariant mass of the 2-jet system produced exclusively  in the reaction~(\ref{eq1}) is shown
on the Fig.~\ref{fig:9}~(a) for the events selected by~(\ref{exp:sel1})\&(\ref{exp:sel3})\&(\ref{exp:sel5}).
Invariant mass of the 2-jet system produced exclusively  in the reaction~(\ref{eq2}) is shown
on the Fig.~\ref{fig:9}~(b) for the events selected by~(\ref{exp:sel2})\&(\ref{exp:sel4})\&(\ref{exp:sel5}).
Efficiency of the signal selection depends on $p_t^{jets}$-cut dominantly. With  selections 
(\ref{exp:sel1})\&(\ref{exp:sel3})\&(\ref{exp:sel5}) we save 2\% of the 
single pion exchange 
events~(\ref{eq1}) and 
9\% of the double pion exchange events~(\ref{eq2}) with (\ref{exp:sel2})\&(\ref{exp:sel4})\&(\ref{exp:sel5}).

\section{Discussions and conclusions}

We propose to measure reactions of single (S$\pi$E) and double (D$\pi$E) pion exchange with 2-jet production 
at the LHC with CMS~\cite{CMS} using the  
ZDC calorimeter~\cite{ZDC}  
to detect leading neutrons. 
Numerical simulation of reactions~(\ref{eq1}) and~(\ref{eq2}) has 
been performed with MONCHER 1.0~\cite{monchermanual} event generator. Background events 
from 2-jet $pp$ inelastic interactions
have been generated by PYTHIA 6.420~\cite{pythia}. In 
this study we investigated effective criteria for selection of events in reactions~(\ref{eq1}) 
and~(\ref{eq2}) and estimated signal/background ratio. On the generator level of simulation the 
perspectives for such measurements look quite positive.

For  2-jet D$\pi$E events we can suppress completely the background from S$\pi$E and $pp$ 
inelastic interactions
using a trigger for neutrons from ZDC (selections~(\ref{exp:sel2})) and properties of the jets measured 
in Barrel, Endcap and HF of the CMS (selection~(\ref{exp:sel4})). The rest of the 2-jet D$\pi$E 
events after all selections, 
9\%, is equivalent to  $\approx 8\div12 \mu$b 
(uncertainty is caused by different predictions for D$\pi$E cross section).

For  2-jet S$\pi$E events we suggested selections (\ref{exp:sel1})\&(\ref{exp:sel3})\&(\ref{exp:sel5}) 
which use only  trigger requirements for neutrons from the ZDC and properties of jets. These selections 
suppress background from 2-jet inelastic events almost completely and save $\approx 2\%$ of the signal, 
which is  equivalent to $\approx 10\div14 \mu$b. 

The data accumulated by the CMS detector (more 
than 300 pb$^{-1}$ at a time of the writing of this text) gives chances to 
extract millions of pure 2-jet S$\pi$E and D$\pi$E events, which are exclusive 
dominantly, for the detailed investigation of PDFs in the pion.

%\underline{\it general conclusions and discussions}
From the theoretical point of view it would be very impressive if we had
parton distributions in the pion in a still unexplored kinematical region, since
pion is a fundamental ``participant'' of the strong interaction. Also comparison
of PDFs in the pion and the proton (anti-proton) can shed light on the mechanism of quark
confinement and differences in the internal (quark, gluon) field structure of 
mesons and baryons.

\section*{Acknowledgements}

This work is supported by the grant RFBR-10-02-00372-a.

\end{document}